# On the origins, growth, and radiative efficiency of J0529-4351, reportedly the fastest-growing known black hole

Yash Aggarwal,

Lamont-Doherty Earth Observatory of Columbia University

Emeritus Associate

Palisades, NY 10965, USA

E-mail: haggarwal@hotmail.com

## ABSTRACT

SMSS J0529-4351 is reportedly the most luminous known quasar, and assuming a standard value of 0.1 for its radiative efficiency $\epsilon$, inferred to be the fastest-growing black hole (BH) accreting ~a solar mass ($M_\odot$)/day. Assessing the implications of this assumption on J0529-4351's seed mass Ms and inception time, we show that the inferred accretion rate $\dot{M}$ is unreasonably high and that $\epsilon$ must be >> 0.1. Then, we derive its $\dot{M}$ and Ms, and for comparison of three other similar-size (~1-2x$10^{10}M_\odot$) BHs at various redshifts, using well-tested empirical scaling relations. The results indicate that J0529-4351 grew from a heavy seed (~2-3x$10^4 M_\odot$), and that its $\dot{M}$ (~10-13x$M_\odot$/yr) is the lowest of the four BHs. However, its radiative efficiency, inferred from its bolometric luminosity and the derived $\dot{M}$, is the highest, which explains why it is the most luminous despite having the lowest $\dot{M}$. This study challenges the prevailing notion that a higher luminosity or a higher Eddington ratio implies a higher $\dot{M}$, highlights the dependence of a BH's luminosity on $\epsilon$, reveals the pitfalls of inferring BH properties assuming a standard value for $\epsilon$, and suggests that the Eddington ratios of high-luminosity BHs may be significantly overestimated.

# 1. INTRODUCTION

Quasar J0529-4351, at a redshift of z=3.962, is powered by a black hole (BH) 17 billion times the mass of the Sun. It is reportedly the most luminous object observed to date and also the fastest-growing known black hole, accreting a solar mass ($M_\odot$) per day (Wolf et al., 2024). Not surprisingly, it has attracted considerable attention from astrophysicists and the media at large, and merits exploration into its origins, growth, and radiative efficiency, which remain obscure. The reported accretion rate is inferred from the black hole's bolometric luminosity, $L_{BOL}$, assuming a standard mean value of ~0.1 for the radiative efficiency $\epsilon$. This mean value, based on the Soltan argument (1982) or its variants by connecting the mass function of local massive black holes with the luminosity function of distant active galactic nuclei (AGNs), is well established (e.g., Yu and Tremane, 2002) and widely used to infer BH properties. A BH's instantaneous radiative efficiency $\epsilon$, however, can differ substantially from the standard value (Shankar et al., 2010), and it has recently been shown to be a function of a BH's mass and redshift (Aggarwal, 2024). Hence, it is imperative to investigate the consequences of this assumption and test the validity of the inferred accretion rate. A simple test is to assess the implications of the inferred accretion rate for the BH's origins (seed mass and inception time). By doing so, we find it implausible that J0529-4351 is growing at a rate anywhere close to one solar mass per day. We show that its radiative efficiency $\epsilon$ must be much higher than the assumed value of 0.1, and that its growth rate is consequently much lower.

We then use well-tested, theoretically supported empirical scaling relations recently reported by Aggarwal (2025) to determine the seed mass Ms and the instantaneous accretion rate $\dot{M}$ for J0529-4351 and three other similar-sized (~1-2×$10^{10} M_\odot$) BHs at various redshifts. Before applying these scaling relations, however, we provide a brief review of their bases and validity. Besides providing estimates of Ms and $\dot{M}$ for each of the four BHs, a comparative analysis of the results reveals similarities in their origins and establishes that J0529-4351 is not the fastest-growing BH observed to date. In addition, we infer each BH's radiative efficiency $\epsilon$ using its reported bolometric luminosity $L_{BOL}$. However, in each case, the inferred $\epsilon$ exceeds the theoretical Kerr limit for a spinning BH. We seek an explanation for this discrepancy and find that different authors have used different correction factors in estimating $L_{BOL}$, Consequently, we re-determine the $L_{BOL}$ of each BH using the reported value of the monochromatic luminosity at 3000 Å and applying a uniform luminosity-dependent correction formula. Thus, redetermining the $L_{BOL}$ and $\epsilon$ for each BH resolves the discrepancy and confirms that J0529-4351 has the highest luminosity and radiative efficiency among the four BHs. We end with a summary of the findings and their broader implications.

# 2. CONTRARY EVIDENCE

A BH's radiative efficiency $\epsilon$ is conventionally defined with respect to the mass inflow rate, and a BH's accretion rate $\dot{M}$ is smaller by a factor of $(1-\epsilon)$, and hence a BH's bolometric luminosity $L_{BOL}$ is defined by Eq.1, where c is the velocity of light.

$$L_{BOL} = \dot{M}c^2 (\epsilon/1-\epsilon) \quad\quad\quad\quad (1)$$

Using two monochromatic luminosities and a standard correction factor, Wolf et al. (2024) determined Log $L_{BOL}$ = 48.37 for J0529-4351, and assuming $\epsilon$=0.1, obtained a mass inflow rate ~413$M_\odot$ / yr that translates into an accretion rate $\dot{M}$~371$M_\odot$/yr or ~ 1$M_\odot$/day. Their best-fit slim disc model yields a similar accretion rate with a radiative efficiency of 0.09. By definition, $L_{BOL}$ is a direct function of a BH's accretion rate $\dot{M}$ and radiative efficiency $\epsilon$ (Eq.1), and assuming $\epsilon$ to be a constant and assigning it a standard value, as is customarily the case, inevitably leads to the conclusion that higher the $L_{BOL}$, the higher the accretion rate. And since the $L_{BOL}$ of this BH is reportedly the highest of any quasar observed to date (Wolf et al., 2024), it inevitably follows that J0529-4351 is the fastest-growing BH observed to date. Evidently, this conclusion is predicated on the assumption that all BHs have similar radiative efficiencies, and the inferred accretion rate for J0529-4351 is based on the unsubstantiated assumption that $\epsilon$ ~0.1. At a sustained rate of 1$M_\odot$/day or close to it, this BH would have accreted the equivalent of its mass of ~17x10$^9$$M_\odot$ in just ~46x10$^6$ years (46Myr), comparable to the well-known Salpeter e-folding time of 45Myr. The implication is that the seed from which this BH grew into its observed mass originated just ~46Myr before the redshift z=3.962 at which it is observed. We can estimate the mass Ms of its seed using the following equation that describes its growth from $t_S$ the origin time of the seed to the age t at which the BH is observed, and where $\delta$ is the duty cycle, $\lambda$ the Eddington ratio, $\epsilon$ the radiative efficiency, and $(t-t_S)$ =46 with time expressed in Myr.

$$M_{BH}(t-t_S) = M_S \exp((1-\epsilon)/\epsilon)\, \lambda\, \delta\, (t-t_S)/450) \quad\quad\quad\quad (2)$$

As discussed later, a BH's $\delta$, $\lambda$, and $\epsilon$ are functions of its redshift and change as a BH evolves. Hence, in Eq.2, their values are averages over a BH's life span $(t-t_S)$ and may differ from their instantaneous values at the redshift at which the BH is observed. However, the inferred lifespan (~46 Myr) of J0529-4351 is rather short, essentially equal to the e-folding time. Hence, we can safely assume the values of $\lambda$ and $\epsilon$ in Eq.2 to be equal to their instantaneous values. Wolf et al. (2024) determined an instantaneous $\lambda$ = 0.9 and assumed $\epsilon$ = 0.1; and the value of the duty cycle $\delta$, or the fraction of time during which the BH is actively accreting, is in this case $\delta$ = 1. Thus, solving Eq. 2, we obtain seed mass Ms ~ 5x10$^{10}$$M_\odot$, which is many orders of magnitude larger than the mass (10$^{4-6}$$M_\odot$) of the heaviest seeds that could conceivably have formed by any of the proposed mechanisms of seed formation (see e.g., review by Volonteri et al., 2021). Furthermore, BH seeds are thought to have formed at z~25-30 (e.g., Barkana and Loeb, 2001; Pacucci and Loeb, 2020; Zubovas and King, 2021), and the recent discovery of the AGN GNZ11 at z=10.6 (Maiolino et al., 2024) supports such an assumption. The clear implication is that the inferred



instantaneous accretion rate is abnormally high. Hence, the assumption that ϵ ~0.1 is erroneous and must be substantially higher.

Shankar et al. (2010) investigated the characteristic values of λ, ϵ, and δ of AGNs at z=3-6. They concluded that to simultaneously reproduce the observed luminosity function and bias, successful models must have $f/\lambda \geq 0.7$, where $f = \epsilon/(1- \epsilon)$. They further stressed that these findings are robust against uncertainties in the obscured fraction of AGNs or in the precise value of the mean bolometric correction. This requirement may not be valid at all redshifts. However, the redshift of J0529-4351 is within the z range of the AGNs inspected by Shankar et al. (2010). And since its Eddington ratio λ determined by Wolf et al. (2024) is 0.9, the instantaneous radiative efficiency presumably ought to be ≥ 0.39 in accord with the foregoing conclusion that ϵ has to be substantially >0.1. Consequently, as per Eq.1, the instantaneous accretion rate should be $\dot{M} \leq 64.5 M_\odot$/yr or ≥ 5.7 times smaller than that inferred assuming =0.1, and the accretion period $(t - t_s)$ ought to be correspondingly much longer.

We can further explore the dependence of a BH's derived property such as its seed mass Ms on the assumptions made concerning the various parameters in Eq.2. First, consistent with the prevailing notion (e.g., Barkana and Loeb, 2001; Pacucci and Loeb, 2020; Zubovas and King, 2021) and substantiation (Aggarwal, 2025) that the seeds of SMBHs formed at or near z=30 when the universe was ~100 Myr old, we can safely assume the same for J0529-4351. In which case $(t-t_s)$ ~1450Myr in Eq.2, using the Hubble constant ($H_\odot$=67.4 km/s/Mpc) and matter density parameter ($\Omega_m$ =0.315) from the Planck group (2020). This lifespan is more than an order of magnitude longer than that derived above, assuming ϵ = 0.1, and we can no longer assume that the average values of λ, ϵ, and δ in Eq. 2 are equal to their instantaneous values. Shankar et al. (2010) found that for AGNs at z =3-6, the duty cycle δ increases as z increases from ~0.2 at z=3.1 to ~0.9 at z=6. Hence, since δ cannot be >1, and since a substantial portion of this AGN's lifespan (~1450Myr) precedes z=6, we can reasonably assume δ=0.9 in Eq. 2 and derive the seed mass Ms for different assumed mean values of λ and ϵ. Thus, using t-ts = 1450 Myr, duty cycle δ = 0.9, and λ = 0.9, Eq. 2 yields a seed mass as low as Ms ~1$M_\odot$ for ϵ =0.1 and as high as Ms ~5x10$^5 M_\odot$ for ϵ=0.2. Alternatively, if we assume average λ=0.45 instead of 0.9, Eq.2 yields Ms ~1.4x10$^5 M_\odot$ for ϵ=0.1 and Ms ~9.2x10$^7 M_\odot$ for ϵ=0.2. These results demonstrate that the derived value of Ms critically depends upon assumed values of λ and ϵ. A mere factor of 2 change in either the input value of ϵ or λ results in many orders of magnitude change in the derived value of seed mass Ms, a finding that further underscores the pitfalls of inferring BH properties based on an assumed value of ϵ.

## 3. SEED MASS, ACCRETION RATE, AND RADIATIVE EFFICIENCY



Fortunately, it turns out that we do not have to assume the values of λ and ε to derive a BH's seed mass Ms using Eq.2. By definition, λ= $L_{BOL}/L_{EDD}$, where the Eddington luminosity $L_{EDD}$ ∝ $M_{BH}$, and as per Eq.1, $L_{BOL}$ ∝ $\dot{M}$ ε/(1- ε). Hence, in Eq. 2, the term $(1 − ε)/ε)δλ$ ∝ $δ(\dot{M}/M_{BH})$, or proportional to the accretion rate per unit BH mass $M_{BH}$. And since a BH's δ decreases as z decreases (Shankar et al. 2010), and its λ decreases and ε increases as z decreases (Aggarwal, 2024), the term $(1 − ε)/ε)δλ$ in Eq.2 or $δ(\dot{M}/M_{BH})$ decreases as z decreases. Thus, defining $δ(\dot{M}/M_{BH})$ ∝ $1/t$, Eq.2 reduces to: $M_{BH}(t) = M_s \exp[β(t- t_s)/t]$, where β is the dimensionless proportionality constant. Note that $M_{BH}$ ∝ Ms and depends much less on the term (t- ts)/t. Hence, BHs with similar $M_{BH}$ and within a narrow age t window should have similar seed masses Ms, assuming ts is constant. Thus, using the mass $M_{BH}$ and age t data for 60 high-z (5.8-7.6) SMBHs with similar $M_{BH}$ and ages within a narrow (~300Myr) time window, Aggarwal (2025) obtained the values of the parameters/constants ts and β by simultaneously solving the 60 equations. The resultant relationship is defined by Eq.3A, where age t is in Myr and $M_{BH}$ and Ms are in solar masses. And using the approximation $1/t ∝ (1+z)^{3/2}$ for high z (Bergström and Goober, 2006), Eq.3A is rewritten (Eq.3B) in terms of redshift z instead of age t.

$$M_{BH} = M_s \exp[14.6(t- 100)/t] \quad (3A)$$

$$M_{BH} = M_s \exp 14.6 [1 - (1 + z)^{3/2} / (1+30)^{3/2}] \quad (3B)$$

See Aggarwal (2025) for greater details as to how these relationships were derived. Note, however, that the only assumption used in deriving Eq.3 is that BH seeds formed at essentially the same epoch (ts=constant), a commonly used assumption in BH growth models (e.g., Pacucci and Loeb, 2020; Zubovas and King, 2021; Fragione and Pacucci, 2023), If, however, the seeds had in fact formed at substantially different epochs, the 60 equations would not have converged and yielded unique values of the parameters β and ts, which a posteriori validates the assumption. Furthermore, note that the derived value of ts (100Myr or z=30) in Eq.3 concurs with that (z=25-30) assumed in numerous studies (e.g., Pacucci and Loeb, 2020; Zubovas and King, 2022; Fragione and Pacucci, 2023; Barkana & Loeb, 2001) and coincides with the timing of the metal-free first massive stars (Couchman and Rees, 1986) that are thought to have collapsed to form stellar-mass seeds (Madau and Rees, 2001; Johnson and. Bromm, 2007) dubbed as Pop III remnants. This relationship was extensively tested by applying it to more than 132,000 SMBHs >$10^6 M_\odot$ at z <2.4 and ~100 at higher redshifts up to z=10.6. The resultant seed masses range from a few solar masses to ≤ $4×10^4 M_\odot$ well within the range of seed masses expected from the different proposed mechanisms of seed formation, and their mass functions concur with the results of simulations of formation of BH seeds (see Aggarwal, 2025). Furthermore, the predictions of Eq.3 were tested against observational data and theoretical considerations. For example, Eq.3 predicts that the maximum size a BH can achieve via luminous accretion amounts to ~$2.2×10^6$Ms, which translates into $(6.6\pm2.2)10^{10} M_\odot$ for the empirically determined upper limit of seed mass Ms = $(3\pm1)×10^4 M_\odot$, in excellent agreement



with the mass of the largest BHs observed to date and a probable theoretical upper limit of ~ $5 \times 10^{10} M_\odot$ proposed by King (2012).

The time derivative of Eq. 3A gives the following relation defining a BH's instantaneous accretion rate Ṁ as a function of its mass $M_{BH}$ and redshift z (see Aggarwal, 2025).

$$\dot{M} (M_\odot/yr) \sim 4.96 \times 10^{12} M_{BH} (1+z)^3 \qquad (4)$$

Note that in the Standard Cosmological Model, the ambient gas density scales as $(1+z)^3$, and hence the accretion rate Ṁ is a direct function of a BH's mass $M_{BH}$ (gravitational reach) and the ambient gas density, consistent with the Bondi (1952) prescription. This relationship was also tested. Perhaps the best illustration of its validity is the following example, which shows that it holds over 13 billion years of cosmic time and ~4 orders of magnitude in BH mass. It predicts accretion rates for BHs in nearby galaxies M87, NGC 3115, and NGC 1600 at z <0.1 in accord with their Bondi accretion rates measured using Chandra X-ray observations. Even more remarkably, Eq.4 predicts a Bondi accretion rate of $>2 \times 10^{-5} M_\odot/yr$ for Sagittarius A*, comparable to its Bondi rate of $\geq 5 \times 10^{-5} M_\odot/yr$ determined by Quataert et al. (1999) using observations of mass-losing stars (see Aggarwal, 2025, and references therein) These concurrences are indeed astonishing because the primary equation (Eq.3), from which Eq.4 is derived, is based on data for BHs at z >5.7 when the universe was less than a billion years old, whereas the Bondi rates for Sagittarius A* and BHs in nearby galaxies are based on observational data in the local universe more than 13 billion years old. Moreover, the high-z BHs used to derive the relationship have $M_{BH} \geq 10^{-9} M_\odot$, whereas Sagittarius A* is ~3 orders of magnitude smaller. Having established the robustness of the foregoing empirical relations, we can confidently apply them to J0529-4351. The other alternative is to use a thin-disc model-based theoretical scaling relation developed by Davis and Laor (2011) and used by Trakhtenbrot et al. (2017) to infer the accretion rate Ṁdisk of 20 high-z (>5.3) BHs. This theoretical relation, however, implies that Ṁdisk is an inverse function of BH mass, which is counterintuitive and conflicts with the conventional notion that the higher a BH's mass, the greater its gravitational reach and, hence, the greater its accretion rate.

We can derive a BH's seed mass Ms using Eq.3, its instantaneous accretion rate Ṁ using Eq.4, and its instantaneous radiative efficiency $\epsilon$ using Eq.1 and the derived Ṁ. Thus, we applied these equations to J0529-4351 and, for comparison, to three other similar-mass AGNs at various redshifts. The results are shown in Table 1. The ages of the four AGNs from the lowest to the highest redshift are respectively ~1551Myr, 1263Myr, 1069Myr, and 870Myr, calculated using the Hubble constant (H☉=67.4 km/s/Mpc) and matter density parameter (Ωm =0.315) from the Planck group (2020) and a flat universe. Note that the only inputs required to derive a BH's Ms and Ṁ are its mass $M_{BH}$ and redshift z, of which the latter is usually precisely known. Hence, the uncertainties in the derived values of Ms and Ṁ are directly proportional to the reported uncertainty in $M_{BH}$, and any uncertainty arising from that of a BH's z is negligibly small. The derived value of radiative efficiency $\epsilon$, on the other hand, is prone to multiple errors arising from the uncertainties in the value of the observed monochromatic luminosity, the correction factor used, and the inferred $\dot{M}$. The uncertainty in $\epsilon$ shown in Table 1 is based on the reported uncertainty in $L_{BOL}$ and that in the derived value of $\dot{M}$. The "real" uncertainty in $\epsilon$, however, is probably much higher. The value of $\epsilon$ shown in parentheses is based on a recalculation of $L_{BOL}$ as discussed later.

**Table 1**

**Derived seed mass Ms, accretion rate $\dot{M}$, and radiative efficiency $\epsilon$**

| Object | Z | $M_{BH}$ ($10^{10} M_\odot$) | $L_{BOL}$ ($10^{48}$ ergs/s) | Ref. | Ms ($10^4 M_\odot$) | $\dot{M}$ ($M_\odot$/yr) | $\epsilon$ |
|---|---|---|---|---|---|---|---|
| J0529-4351 | 3.962 | 1.7±0.12 | 1.86±0.22 | 1 | 2±0.14 | 10.34±0.7 | 0.75±0.03 (0.48±0.03) |
| J2167-3602 | 4.692 | 2.14±0.38 | 1.48±0.57 | 2 | 3.1±0.55 | 19.6±3.5 | 0.57±0.12 (0.29±0.03) |
| J0306+1853 | 5.363 | 1.07±0.27 | 1.32±0.08 | 3 | 1.9+0.55 | 13.7±3.5 | 0.63±0.07 (0.29±0.04) |
| J0100-282 | 6.3 | 1.24±0.19 | 1.62±0.47 | 4 | 3.0±0.46 | 23.9±3.5 | 0.54±0.10 (0.26±0.02) |

$M_{BH}$ = BH mass; $M_\odot$ = Solar mass; $L_{BOL}$ = Bolometric luminosity. The radiative efficiency $\epsilon$, not in parentheses, is based on the reported $L_{BOL}$ in column 4; the one in parentheses is based on revised $L_{BOL}$ determined using monochromatic luminosity at 3000 Å and the correction formula from Netzer (2019). References: 1 = Wolf et al. (2024); 2 = Lai et al. (2023); 3 = Wang et al. (2015); and 4 = Wu et al. (2015).

The four AGNs in Table 1 have similar masses $M_{BH}$ (within ~40% of $1.5 \times 10^{10} M_\odot$) and so do their seed masses Ms (within ~30% of $2.5 \times 10^4 M_\odot$), in Table 1. This concordance between $M_{BH}$ and Ms is not surprising since $M_{BH}$ is a direct function of Ms and depends to a much lesser extent on the BH's age t or redshifts z (see Eq.3). Note that all four BHs have heavy seeds, whose masses concur with and are well within the limits ($10^{4\text{-}5} M_\odot$) of seeds envisioned to have formed by the direct-collapse mechanism (Bromm and Loeb, 2003; Begeleman et al., 2006; Lodato and Natarajan, 2006; Shang et al., 2010) dubbed as DCBHs, or by hierarchical growth of BHs in dense stellar clusters (Davies et al., 2011). These are among the most massive BHs observed to date, and their inferred Ms are within the empirical upper limit of $(3\pm1) \times 10^4 M_\odot$ determined by Aggarwal (2025). We conclude that J0529-4351 originated from a heavy seed of $\sim 2 \times 10^4 M_\odot$ formed or assembled at z~30 when the universe was ~ 100 Myr old.

Of the four AGNs, the accretion rate $\dot{M}$ of J0529-4351 is clearly not the highest. On the contrary, it is ~2 times lower than that of J0100-282, significantly lower than that of J2167-3602, and probably lower than that of J0306+1853. Note, however, that the $M_{BH}$ of J0529-4351 may be underestimated by a factor of ~1.26 as discussed later. Even then, its revised $\dot{M}$ (~13$M_\odot$/yr) would be lower than those of J0100-282 and J2167-3602, and comparable to that of J0306+1853. And if there are any systematic errors in determining BH masses, they should equally affect all four BHs. We conclude, therefore, that J0529-4351 is not the fastest-growing BH observed to date. Furthermore, its instantaneous accretion rate of ~10-13$M_\odot$/yr derived here is orders of magnitude lower than that inferred by Wolf et al. (2024) assuming $\epsilon=0.1$.

Of the four AGNs, the bolometric luminosity $L_{BOL}$ of J0529-4351 is the highest (ignoring the uncertainty in the $L_{BOL}$ of J0100-282) in accord with the Wolf et al. (2024) finding that it is the most luminous quasar observed to date. However, the values of $\epsilon$, derived using the reported $L_{BOL}$ of each BH and its inferred $\dot{M}$ in Table 1, exceed the Kerr limit of ~42.3% (Bardeen, 1970) for the radiative efficiency of a maximally spinning BH in an ideal thin accretion disc. We explored the reasons for this apparent discrepancy. A BH's $L_{BOL}$ is prone to multiple sources of errors (see, e.g., Runnoe et al. 2012), of which the uncertainty in the bolometric correction factor is probably the most important. Briefly, a BH's $L_{BOL}$ is not measured directly but is almost universally estimated by observing a BH at a few specific wavelengths and by applying a bolometric correction factor specific to the wavelength. Different methods for making these corrections can lead to significant uncertainties and large errors. In fact, the correction factors proposed by different authors for different wavelengths (e.g., Elvis et al. 1994; Richards et al. 2006; Nemmen & Brotherton, 2010; Runnoe et al. 2012) differ considerably. And in the linear correction formulae recommended by Runnoe et al. (2012), the non-zero intercept has very large uncertainties, resulting in more than an order of magnitude uncertainty in $L_{BOL}$. Moreover, Trakhtenbrot and Netzer (2012) and Netzer (2019) found that the correction factor is apparently a function of luminosity and decreases as the BH's luminosity increases, and the constant correction factors often used result in overestimates of $L_{BOL}$, especially for BHs with high luminosities.

Hence, we examined how the reported $L_{BOL}$ for each of the four BHs was calculated.. We found that one of the monochromatic luminosities used was at 3000 Å, but different authors applied different correction factors, and none accounted for the dependence of the correction factor on the magnitude of the monochromatic luminosity. We recalculated the $L_{BOL}$ of each of the four BHs using the reported value of the luminosity at 3000 Å and applying the correction formula for this wavelength defined in Netzer (2019), along with a uniform 0.75 anisotropy factor from Runnoe et al. (2012), also used by Wolf et al. (2924). Note that Netzer's bolometric correction formula is based on thin-disk accretion, and so is the theoretical Kerr limit. The correction factors in Netzer, however, are for a disk inclination or a viewing angle of ~56 degrees. And if the assumed viewing angle is lower, the applicable correction factor is correspondingly lower (see Netzer, 2019). The revised $L_{BOL}$ values are systematically lower by factors of ~3-4. They are respectively: $5.4 \times 10^{47}$; $4.5 \times 10^{47}$; $3.2 \times 10^{47}$; and $4.68 \times 10^{47}$ for J0529-4351, J2167-3602, J0306+1853, and J0100-282 in ergs/s. The revised $L_{BOL}$ for J0529-4351 is still the highest of the four BHs, but its revised Eddington ratio λ is ~0.25, comparable to that (0.16-0.29) of the other three BHs, but much lower than the ~0.9 obtained by Wolf et al. (2024). These recalculations of $L_{BOL}$ and λ suggest that the reported $L_{BOL}$ and λ of other BHs calculated using constant correction factors may also be significantly overestimated, especially for BHs with high luminosities.

The resulting value of ϵ for each of the four BHs, using its revised $L_{BOL}$ and the derived $\dot{M}$, is shown in parentheses in the last column in Table 1. The revised results show that, except for J0529-4351, the ϵ for the three other BHs are well below the Kerr limit of ~42.3%. Moreover, the revised results show that the lower the BH's redshift, the higher the ϵ, in accord with the recent finding (Aggarwal, 2024) that ϵ is an inverse function of z. The revised ϵ for J0529-4351 is, however, still above the Kerr limit, albeit by only ~13%. There is at least one other important factor that could result in an overestimation of its ϵ, namely, an underestimation of its mass. There are indications that its mass is likely $> 17 \times 10^{10} M_\odot$. The estimates of its mass using MgII and CIV lines range from $~1-28 \times 10^{10} M_\odot$, and from its continuum shape, Wolf et al. (2024) found $15-28 \times 10^{10} M_\odot$ with a median value ~26% higher than the adopted value of $17 \times 10^{10} M_\odot$. In which case, its derived accretion rate would be ~26% higher than that in Table 1, and the revised ϵ of ~0.42 would comply with the Kerr limit, albeit near the upper limit. Note that the revised estimates of the derived ϵ do not take into account the uncertainty in the bolometric correction factor, which, as per Netzer (2019), is difficult to ascertain. What is clear, however, is that J0306+1853 has the highest ϵ, which explains why it is the most luminous of the four BHs despite having the lowest accretion rate.

## 4. CONCLUSIONS

Wolf et al.'s (2024) conclusion that J0529-4351 is the fastest growing black hole, accreting a solar mass per day, is predicated on the assumption that all BHs have the same standard radiative efficiency ϵ ~0.1. Assessing the implications of this conclusion on the origins of J0529-4351, we showed that the reported instantaneous accretion rate $\dot{M}$ is unreasonably high and that ϵ must be

>> 0.1. Moreover, we showed in general that the derived value of a BH's property, such as the mass Ms of its seed, is highly sensitive to the assumed inputs of $\epsilon$ and/or the Eddington ratio, resulting in highly erroneous inferences.

Using well-tested empirical scaling relations, we determined the Ms and $\dot{M}$ of this and three other similar-size BHs at various redshifts. We infer that J0529-4351 originated from a heavy seed (~2-3x$10^4 M_\odot$) and that its $\dot{M}$ (~10-13$M_\odot$/yr) is the second lowest, if not the lowest, of the four BHs. Using the derived $\dot{M}$ and the reported bolometric luminosity $L_{BOL}$, we obtained $\epsilon$ of each BH. Finding that the resulting $\epsilon$ exceeded the theoretical Kerr limit of ~42.3%, we examined the basis for the reported $L_{BOL}$ for each BH. Finding that different authors used different bolometric correction factors, we recalculated each $L_{BOL}$ using the BH's monochromatic luminosity at 3000 Å and uniformly applying the relevant luminosity-dependent correction formula from Netzer (2019). We found that the reported $L_{BOL}$ were systematically overestimated by factors of 3-4 and that the revised Eddington ratio of J0529-4351 is ~0.25. Applying the revised $L_{BOL}$, the discrepancy between the derived $\epsilon$ and the theoretical Kerr limit is resolved. Moreover, the results show that J0529-4351 has the highest radiative efficiency, which explains why it is the most luminous despite having the lowest accretion rate.

The finding that the $\dot{M}$ of J0529-4351 is the lowest and its $\epsilon$ the highest is consistent with and explainable by the following observations. The accretion rate $\dot{M}$ is a direct function of a BH's mass and the ambient gas density, of which the latter scales as $(1+z)^3$. Hence, of the BHs with similar masses, the one at the lowest z should have the lowest $\dot{M}$. Of the four AGNs, J0529-4351 is at the lowest z, and its $\dot{M}$ is the lowest. On the other hand, using data for groups of tens of thousands of BHs, Aggarwal (2024) found that a BH's $\epsilon$ is an inverse function of its $M_{BH}$ and redshift z. Hence, for BHs with similar $M_{BH}$, the one at the lowest z should have the highest $\epsilon$. Of the four AGNs, J0529-4351 is at the lowest z and its derived $\epsilon$ is the highest.

The results of this study challenge the widespread notion that higher luminosity or Eddington ratio implies higher accretion rate. The results reveal that J0529-4351 is the most luminous of the four BHs, not because it is the fastest-growing, but because its radiative efficiency is the highest - a finding that highlights the critical role, heretofore ignored, that $\epsilon$ plays in determining a BH's luminosity. This study, in fact, reveals the pitfalls of inferring BH properties assuming a standard value for $\epsilon$. And finally, this study suggests that the reported $L_{BOL}$ and Eddington ratios based on constant bolometric correction factors that do not consider a BH's luminosity may be systematically overestimated, especially for high-luminosity BHs.

**ACKNOWLEDGMENTS**: I thank Manuel Chirouze for constructive criticism.

**DATA AVA I L A B I L I T Y**: No new data were generated in this study.

**ETHICS DECLARATION**: The authors declare no competing interests.